\documentclass{article}
\usepackage{amssymb}
\usepackage{amsmath}
\usepackage{graphicx}
\begin{document}

\title{Cosmological evolution with a logarithmic correction in the dark
energy entropy}
\date{ }
\author{C. Barbachoux, J. Gariel and G. Le Denmat \\
LERMA, UMR/CNRS 8112 \\
ERGA Universit\'{e} P6, bc142, \\
4 place Jussieu, 75252 Paris France}
\maketitle

\begin{abstract}
In a thermodynamical model of cosmological FLRW event horizons for
the dark energy (DE), we consider a logarithmic corrective term in
the entropy to which corresponds a new term in the DE density. This
model of $\Lambda (t)$ in an interacting two-component cosmology
with cold dark matter (DM) as second component leads to a system of
coupled equations, yielding, after numerical resolution, the
evolutions of $\Lambda (t)$, the Hubble $H(t)$, vacuum density
$\Omega _{\Lambda }(t)$, deceleration $q(t)$ and statefinder $R(t)$
and $S(t)$ parameters. Its results, compatible with an initial
inflation and the current observations of the so-called
``concordance model", predict a graceful exit of early inflation and
the present acceleration and solve in the same time the age and
coincidence problems. Moreover they account for the low-$l$ CMBR
power spectrum suppression.

\textbf{keywords} : Dark energy theory, SnIa constraints, CMBR (low
multipoles) theory.
\end{abstract}

\section{Introduction}

The observations of SNe of type Ia \cite{article3}\cite{article4},
showing that the universe has probably recently entered on a phase
of acceleration, raised up a renewal of interest in the cosmological
constant (CC) problem \cite{article1} \cite{article2}.

Since then an increasing number of observational evidences for such
an acceleration has been settled. This acceleration together with
other cosmological indicators such as the flatness ($k=0$) from the
CMB anisotropies observed by Boomerang, Proxima and WMAP
\cite{article5}, the present values $H_{0}$ by the HST key project
\cite{article6}, $\Omega _{b0}$ and $\Omega _{m0}$ by observations
of the light elements, SBBN and cluster baryon fraction from X-ray
emission \cite{article7}), led to the so-called ``concordance model"
\cite{article8} based on the $\Lambda$CDM model. This model is the
simplest improvement of the old standard cosmological model
(including inflation). More precisely, the simplest theoretical
candidate explaining this accelerated expansion is a cosmological
constant $\Lambda $, or some component, called dark energy (DE): any
term in the energy-momentum tensor which can produce the same
negative pressure, such as a scalar field \cite{article9}, called
quintessence \cite{article9b}, or a bulk viscosity \cite{article10}.

However, the re-introduction of the CC $\Lambda $ leads to new
intriguing puzzles. Particularly, we have to face the old CC
problem: the discrepancy between the predictions of the effective
Quantum fields theory concerning the large vacuum energy density and
the tiny upper limits observed in astrophysics for the CC. Besides,
it gives rise to a new issue, the so-called ``coincidence problem",
where one has to explain why the CC starts recently to dominate the
matter. Assuming $\Lambda =constant$, it can
only be solved by an unsatisfying fine-tuning. Finally, the introduction of $%
\Lambda $ needs also to be conciliated with the inflation in the
early universe.

In these conditions, we have to improve the $\Lambda CDM$ model in order to
solve these puzzling issues.

Therefore, we propose a $\Lambda (t)CDM$ model, namely the simplest
modification of the $\Lambda CDM$ model which allows the best
account of the theoretical expectations and the observed data. Some
``ansatze" were already proposed to solve the CC problem before the
discovery of the present universe acceleration (for a review, see
\cite{article11}). By contrast, the present model, namely the event
horizon Thermodynamics (ehT) model, is derived from a
thermodynamical approach and not based on an ad hoc hypothesis for
the law $\Lambda(t)$. Such a phenomenological model is an
indispensable preliminary to outline the main laws that any further
treatment at the statistical or/and quantum level will have to
reproduce, as for example the Stefan's law has guided the Planck
law's advent for the black body radiation. We thus proposed a model
with a variable vacuum energy density
\cite{article13}\cite{article27}, based on the generalization of the
Thermodynamics of the black hole event horizons (BH
e.h.)\cite{article19} to the
Friedmann-Lema\^{\i}tre-Robertson-Walker event horizons (FLRW
e.h.)\cite{article12}. This model allows to reconstitute the entire
evolution from the early inflation to late acceleration (for another
approach, see also \cite{article48}).

In this article, we extend this phenomenological model by introducing a
logarithmic correction in the $\Lambda$-entropy. Such a corrective term in
the entropy was recently computed for the entropy of any thermodynamical
system and applied to Black Hole by S.Das et al. \cite{article14}. This
correction due to statistical fluctuations around an equilibrium state has
an universal character. As the cosmological FLRW e.h. presents notable
thermodynamical analogies with the BH e.h. (albeit some differences \cite%
{article15}), we examine the effects of the introduction of this new
term in the $\Lambda (t)CDM$ model. It appears that the derived
expression for $\Lambda (r)$, with $r$ the radius of the event
horizon (e.h.), can then be naturally interpreted as a
renormalisation relation and allows to describe the main observed
features of the evolution of the universe. In this scheme, the
obtained universe is initially submitted to a local phase of de
Sitter inflation ($\Omega _{\Lambda }(0)=1$). The vacuum density
parameter begins to decrease and there is a graceful exit of
inflation. The parameter $\Omega _{\Lambda }$ passes through a small
minimum, then grows again, which allows to solve the ``co{i}ncidence
problem". It tends
asymptotically to a de Sitter universe. At any instant $t$, $H(t)$ and $%
\Lambda (t)$ are decreasing functions.

The paper is organized as follows. In Section 2, we set the basis of
the general relativistic framework in which our model for $\Lambda
(t)$ will be developed. In Section 3, we present the ehT model for
the vacuum (or DE) energy density. Then in Section 4, we examine the
influence of the entropy's correction term in the vacuum energy
density and its relevance in the renormalisation approach. In
Section 5, we deduce the equations of evolution for $H(t)$ and
$r(t)$, and present in section 6, their numerical resolution, in
particular, the graphic of $\Omega _{\Lambda }(t)$. Section 7 is
devoted to a brief conclusive summary.

\section{\protect\bigskip An interacting two-component model}

Let us briefly recall the minimal assumptions underlying our model.

In order to set our notations, let us first write the Einstein equations:%
\begin{equation}
R_{\alpha \beta }-\frac{R}{2}g_{\alpha \beta }+\Lambda _{0}g_{\alpha \beta
}=\chi T_{\alpha \beta }\text{.}
\label{eq1}
\end{equation}
with $\chi =\frac{8\pi G}{c^{4}}$ and $\Lambda _{0}$ a positive
constant (the GR bare cosmological constant). Only for this choice
of coefficients in Eq. (\ref{eq1}), the Bianchi identity $\nabla _{\beta
}T^{\alpha \beta }=0$ is satisfied
\cite{article16}\cite{article17}.

Assuming a perfect fluid of energy density $\rho_T$ and pressure
$P_T$, the energy-momentum tensor is given by  $T_{\alpha \beta
}=\rho _{T}u_{\alpha }u_{\beta }-P_{T}\Delta _{\alpha \beta }$
with $\Delta _{\alpha \beta }:=g_{\alpha \beta }-u_{\alpha
}u_{\beta }$, and the index "$T$" means "Total".

Following \cite{article13} and \cite{article27}, the perfect fluid
is supposed to
admit two components: $\rho _{T}=\rho _{1}+\rho _{2}$ and $P_{T}=P_{1}+P_{2}$%
.

The first one corresponds to vacuum, i.e., it obeys the
``pseudo-barotropic" equation of state : $P_{1}=\omega \rho _{1}$
with $\omega=-1 $.

The second component obeys a barotropic equation of state given by $%
P_{2}=(\gamma -1)\rho _{2}$. In section 6, we consider dust
($\gamma=1$).

In accordance with the WMAP observational results of the CMB's
anisotropies \cite{article5}, the metrics is taken to be the
spatially flat ($k=0$) FLRW metrics. With these assumptions,
Equation (\ref{eq1}) reduces to
\begin{equation}
R_{\alpha \beta }-\frac{R}{2}g_{\alpha \beta }=\Lambda g_{\alpha \beta
}+\chi T_{2\alpha \beta }
\label{eq2}
\end{equation}
with $T_{2\alpha \beta }:=\rho _{2}u_{\alpha \beta }-P_{2}\Delta _{\alpha
\beta }$ and%
\begin{equation}
\Lambda :=\Lambda _{vac}-\Lambda _{0}.
\label{eq3}
\end{equation}%
As $\Lambda _{vac}:=\chi \rho _{1}$ with $\rho _{1}$ variable, $\Lambda $
varies. In the following, we refer to $\Lambda$ as the `` variable
cosmological constant". It can be considered as an ``effective" CC. The
corresponding dark energy (DE)-component has the energy density : $\rho _{\Lambda }=%
\frac{\Lambda }{\chi }$.

We do not make any further assumption on a separated energy
conservation for each component. In the old standard cosmological
model, this assumption is made only for each era of ``domination"
of one component (e.g. ``era dominated by the radiation", ``era
dominated by the matter", today ``era dominated by the DE", or in
the very early universe ``era dominated by the inflation (or the
inflaton)"). If such an assumption can seem reasonable for these
domination eras (and so appears more as an approximation), it is
not the case for the transient periods, as for instance the very
recent epoch of transition from dark matter (DM) to DE domination.

Anyway, the complete energy conservation due to the Bianchi identity
is more general than the separated conservations, which necessitate
some supplementary assumption. For example, in the $\Lambda (t)$ and
quintessence models, a separated conservation law for each component
implies a ``variable" equation of state. By contrast, the present
model assumes the equations of state in a thermodynamical sense,
i.e. given once for all, with an interaction between the two
components. This kind of approach with
components interaction has already been considered (see e.g. \cite{article18}%
\cite{article18b}).

Let us sketch the rough scenario of the universe evolution. Initially,
setting $\rho _{1}=\rho _{Max}$ and $\rho _{2}=\rho _{2\min }\simeq 0$, the
universe undergoes a de Sitter inflation, with $\Lambda _{M}=\Lambda
_{vacM}-\Lambda _{0}> 0$. Then $\rho _{1}$ decreases or equivalently, $%
\Lambda$ decreases with the universe expansion. In the same time,
$\rho _{2} $ is increasing. Finally, $\rho _{1}$ tends to $\Lambda
_{0}$ or equivalently $\Lambda $ tends to $\Lambda _{\min }\simeq 0$
,which is again a de Sitter evolution. Today, we observe a very
small value for $\Lambda =\chi \rho _{1}-\Lambda _{0}$ with a very
large vacuum energy density $\rho
_{1}$ given by quantum field theory (today $\rho _{1}$ is of the order of $%
\Lambda _{0}$). For a reasonable (Planck) largest cut-off, we obtain $\frac{%
\Lambda }{\chi \rho _{1}}=1-\frac{\Lambda _{0}}{\chi \rho _{1}}\sim
10^{-120} $. Assuming a variable cosmological ``constant" solution
(as considered here), this scheme appears as an hardly intelligible
fine-tuning. It is the so-called ``old problem" of the cosmological
constant (CC). Besides, the recent domination of the vacuum energy
by dust constitutes the ``new puzzle" of the CC (or ``coincidence
problem''). At the beginning, $\rho_2$ represents the energy density
of the radiation and the ultrarelativistic matter. For sake of
simplicity, we shall only consider in the following (section 6) the
non relativistic matter ($\gamma=1$).

We propose a phenomenological model of this DE or, equivalently,
this variable cosmological ``constant" (or this vacuum energy
density) able to produce the main properties of this scenario. More
particularly, the two problems of the CC are solved in the framework
of our model where $\Lambda$ runs from a large initial value
$\Lambda _{M}$ (a huge inflation in the very early universe) to a
weak final constant $\Lambda _{m}$ which yet starts again to
dominate, leading to ``the coincidence problem''. In the two cases,
the same DE, with a density $\Lambda $ is used to explain the
universe acceleration. This DE is negligible or dominates, depending
on the considered period, and its density decreases all along the
universe expansion.

\section{Thermodynamical model of the cosmological FLRW e.h.}

The variable DE density $\Lambda =\chi \rho _{\Lambda }$ is the first
component of the two perfect fluids constitutive of the universe. In the de
Sitter space-time, this cosmological ``constant", or vacuum energy density,
obeys the equations of state \cite{article19}\cite{article20}:
\begin{equation}
\Lambda +\chi P_{\Lambda }=0
\label{eq4}
\end{equation}%
\begin{equation}
\Lambda =\frac{3k^{2}}{c^{2}\text{%
h{\hskip-.2em}\llap{\protect\rule[1.1ex]{.325em}{.1ex}}{\hskip.2em}%
}^{2}}T_{\Lambda }^{2}
\label{eq5}
\end{equation}
where $P_{\Lambda }$ and $T_{\Lambda }$ represent its (negative) pressure
and its temperature respectively,
h{\hskip-.2em}\llap{\protect\rule[1.1ex]{.325em}{.1ex}}{\hskip.2em}
is the Planck constant, $c$ the light velocity and $k$ the Botlzmann
constant. Each thermodynamical variable is linked to the proper radius $r$
of the event horizon at a given instant by the relations

\begin{equation}
T_{\Lambda }=\frac{\text{%
h{\hskip-.2em}\llap{\protect\rule[1.1ex]{.325em}{.1ex}}{\hskip.2em}%
}c}{kr}\text{ ,}
\label{eq6}
\end{equation}%
\begin{equation}
\Lambda =\frac{3}{r^{2}}\text{ .}
\label{eq7}
\end{equation}
We can also introduce the horizons number density $n_{\Lambda }$
defined as usual in Thermodynamics by the inverse of the specific
volume of
the horizon \cite{article13}%
\begin{equation}
n_{\Lambda }=\frac{3}{4\pi r^{3}}\text{ .}
\label{eq8}
\end{equation}
Introducing the expression (\ref{eq6}) of the temperature in the preceding relation,
we obtain a third equation of state:

\begin{equation}
n_{\Lambda }=\frac{3}{4\pi }(\frac{k}{\text{%
h{\hskip-.2em}\llap{\protect\rule[1.1ex]{.325em}{.1ex}}{\hskip.2em}%
}c})^{3}T_{\Lambda }^{3}=\frac{\Lambda ^{\frac{3}{2}}}{4\pi \sqrt{3}}\text{.}
\label{eq9}
\end{equation}

Equation (\ref{eq9}) together with Eq. (\ref{eq4}) and (\ref{eq5})
define a local equilibrium state for the DE component in the de
Sitter spacetime. Let us now consider the same DE component in the
spatially flat FLRW space-time. The three thermodynamical equations
of state (\ref{eq4}), (\ref{eq5}) and (\ref{eq9}) have to remain
valid in FLRW spacetime because the thermodynamical equations of
state of any actual component (e.g. $P=\frac{\rho }{3}$, or the
Stefan law $\rho =\sigma T^{4}$, for the equilibrium radiation) are
always supposed to be independent of the chosen spacetime. Besides,
the definition (\ref{eq8}) of $n_{\Lambda }$ remains also valid in
the (spatially flat) FLRW spacetime. Consequently, from (\ref{eq8})
and (\ref{eq9}), we find that this DE component is associated to the
FLRW e.h. via the relation (\ref{eq7}). Using (\ref{eq7}),
(\ref{eq8}) and (\ref{eq9}), we obtain the expression (\ref{eq6})
for $T_\Lambda$ and (\ref{eq5}) for $\Lambda$ in the FLRW spacetime
too.

Relation (\ref{eq7}) has also been independently derived from
holographic principle \cite{article21}. Albeit the choice of $r$ is
not necessarily \textit{a priori} given in the holographic approach
(e.g. the Hubble horizon $H^{-1}$ can be
chosen instead \cite{article22}\cite{article23}, or the particle horizon $%
r_{p}$ as well \cite{article24}), it is the unique choice in this framework
that leads to an admissible equation of state \cite{article25} and which is
strongly supported by the SnIa observational results \cite{article26}\cite%
{article27}.

Differentiating the definition of $r$ with respect to the time (e.g. see
eqs.(3.4) and (3.6) in \cite{article13} or (9) and (10) in \cite{article27}%
)) yields the relation%
\begin{equation}
H-\frac{\overset{\cdot }{r}}{r}=\frac{c}{r}
\label{eq10}
\end{equation}
where $H$ is the Hubble parameter and the dot means the time derivative.

In order to set a complete local thermodynamics, we need to
determine the entropy density $ns_\Lambda$ where $s_\Lambda$ is the
specific entropy. To reach this goal, we use the Gibbs equation
which at the specific level is given by:
\begin{equation}
T_{\Lambda
}ds_{\Lambda }=d\varepsilon _{\Lambda }+P_{\Lambda }dv_{\Lambda }
\label{eq10bis}
\end{equation}
where $\varepsilon _{\Lambda }\equiv \displaystyle\frac{\Lambda
}{n_{\Lambda }\chi }$ is the specific energy, $v_{\Lambda }\equiv
\frac{1}{n_{\Lambda }}$ the specific volume. With Eq. (\ref{eq4}),
we deduce
\begin{equation}
\chi n_{\Lambda }T_{\Lambda }d{s_{\Lambda }}=d\Lambda \text{.}
\label{eq11}
\end{equation}

\section{Fluctuations and renormalization}

In a recent article, a logarithmic corrective term in the expression of the
entropy of any thermodynamical system was derived by S.Das et al. \cite%
{article14}. This correction due to small thermal fluctuations around an
equilibrium reads for BH entropy:

\begin{equation}
s=\pi r^{2}-\alpha ^{\prime }\ln \pi r^{2}
\label{eq12}
\end{equation}

where $\alpha ^{\prime }$ is a constant. Introducing the same correction in
the cosmological eh model, the expression of the specific entropy becomes
\begin{equation}
s_{\Lambda }=-4(\frac{k}{L_{P}^{2}})(\pi r^{2}+2\pi \alpha \ln \pi r^{2})+C,
\label{eq13}
\end{equation}
where $L_{P}$ is the Planck length (defined by $L_{P}^{2}:=\chi $%
h{\hskip-.2em}\llap{\protect\rule[1.1ex]{.325em}{.1ex}}{\hskip.2em}%
$c=4l_{P}^{2}$, with $l_{P}^{2}:=\frac{Gh}{c^{3}}$ in the I.S. units), $%
\alpha $ is a constant of dimension $[r^2]$ and $C$ an arbitrary constant. As usual, the sign of the
entropy is negative for the cosmological horizon (e.g. \cite{article12} \cite%
{article13}). As noticed by S.Das et al. \cite%
{article14} and Major and Setter \cite{article28}, when there is not
conservation of the number of particles in an open sytem, the sign
``-'' before the $\alpha $-term has to be replaced by a sign ``+''.
This is precisely the case here for our $\Lambda$-component, which
is not separately conserved.

Let us split the constant $C$ into
\begin{equation}
C=4\pi \frac{k}{L_{P}^{2}}(2\alpha \ln \pi r_{2}^{2}+r_{3}^{2}\text{ ),}
\label{eq14}
\end{equation}
where $r_2$ is associated to the corrective term and $r_3$ to the expression
of $s_\Lambda$ without correction. The expression of  $s_{\Lambda }$ becomes
\begin{equation}
s_{\Lambda }=8\pi \frac{k}{L_{P}^{2}}\alpha (\frac{r_{3}^{2}-r^{2}}{%
2r_{1}^{2}}-\ln \frac{r^{2}}{r_{2}^{2}})\text{.}
\label{eq15}
\end{equation}

The constant factor $\alpha :=r_{1}^{2}$ has the dimensions of a surface.
Let us remark the omission of the Bekenstein factor $\frac{k}{l_{P}^{2}}$ in
most papers by choice of the Planck units ($h=c=k=G=1$). The expressions
of $r_2$ and $r_3$ are constrained by setting
\begin{equation}
\left\vert \frac{r_{3}^{2}-r^{2}}{2r_{1}^{2}}\right\vert \gg \left\vert \ln
\frac{r^{2}}{r_{2}^{2}}\right\vert \text{ , }\forall r\geq r(0)
\label{eq16}
\end{equation}
where $r(0)$ is the initial event horizon (see Section $6$). The simplest
choice for $r_{2}$ and $r_{3}$ is given by $r_{2}=r_{3}=r_{M}$ , where $r_{M}$
is the maximum radius (possibly infinite). Hence, $s_{\Lambda }$ is always
positive with $s_{\Lambda }=0$ when $r=r_{M}$. For $r=r_{M}-\epsilon $ with $%
\epsilon \ll r_{M}$, we deduce $r_{3}^{2}-r^{2}\simeq 2\epsilon r_{M}$ and $%
2r_{1}^{2}\ln (\frac{r_{M}^{2}(1-2\frac{\epsilon }{r_{M}})}{r_{M}^{2}}%
)\simeq -2r_{1}^{2}\frac{2\epsilon }{r_{M}}$. A corrective term is only
obtained if $r_{1}\ll r_{M}$ as will be verified in Section 6. Finally, for $%
r=r_{m}$, we have $(r_{M}^{2}-r_{m}^{2})-2r_{1}^{2}\ln ((\frac{r_{m}}{r_{M}}%
)^{2})\simeq r_{M}^{2}+4r_{1}^{2}\ln (\frac{r_{M}}{r_{m}})$. For $r_{1}\ll
r_{M}$ , the $\ln $-term is a corrective term and tends to $s\rightarrow
s_{M}=\frac{8\pi k}{L_{P}^{2}}r_{M}^{2}$. It follows that $r_{1} $ has not
necessarily to be very small with respect to $r$ but only with respect to $%
r_M$.

Let us examine the effects of such a new term in $s_{\Lambda }$ (\ref{eq13}) on the expression of $\Lambda $ .

Differentiating Eq.(\ref{eq13}) for $s_\Lambda$ with respect to $r$ and
introducing this expression into Equation (\ref{eq11}) with (\ref{eq6}) and (\ref{eq8}), we obtain a
differential equation for $\Lambda $ and $r$, which yields after integration
\begin{equation}
\Lambda =\frac{3}{r^{2}}(1+(\frac{r_{1}}{r})^{2})\text{ , }r_{1}^{2}:=\alpha
\geq 0\text{,}
\label{eq17}
\end{equation}
where the boundary conditions are set such that $\Lambda$ tends to $0$ when $%
r\rightarrow \infty $. This relation is an extension of the
expression (\ref{eq7}) of $\Lambda$ when thermodynamical fluctuations are taken into
account.

It is worth noting that this relation is strongly akin to a renormalization
relation (e.g. see first Eq.(3.3) in \cite{article29}, or Eq. (33) in \cite%
{article30}, see also \cite{article33b}). In the same way, Padmanabhan (\cite%
{article31}, Eq.(19)) and Myung (\cite{article32}, Eq. (\ref{eq13})) write a
renormalized vacuum energy density by stressing a hierarchical structure of
the form:
\begin{equation}
\rho _{vac}=\frac{3\text{%
h{\hskip-.2em}\llap{\protect\rule[1.1ex]{.325em}{.1ex}}{\hskip.2em}%
}c}{L_{1}^{2}L_P^2}[1+(\frac{L_{1}}{L})^{2}+(\frac{L_{1}}{L})^{4}+...]\text{,%
}
\label{eq18}
\end{equation}
where the first term is interpreted as the energy density necessitating to be
renormalized, the second term as the vacuum fluctuations and the third
one as the de Sitter thermal energy density (the dimensional constants and
the factor $3$ were restored for convenience). Padmanabhan and Myung have
introduced this form in the case $L_1=L_P$, here we use an intermediate
scale $L_1$ greater than $L_P$. A value of $L_1$ accurate to observational
results will be determined in Section 6.  Setting $\Lambda _{1}:=\frac{3}{%
L_{1}^{2}}$, and multiplying the relation (\ref{eq18}) by $\chi$, we obtain:
\begin{equation}
\Lambda _{vac}-\Lambda _{1}=\frac{3}{L^{2}}[1+(\frac{L_{1}}{L})^{2}+...]%
\text{.}
\label{eq19}
\end{equation}
Assuming $\Lambda_1$ to be the bare CC $\Lambda_0$ of the G.R. (eqs.
(\ref{eq1}) and (\ref{eq3})), this expression is of the same form as
the expression (\ref{eq17}) derived in our model. Let us discuss an
intriguing point. The second term in the r.h.s. of (\ref{eq17}) has
by construction to be interpreted as a term of fluctuations, whereas
it is the first one in the r.h.s. of (\ref{eq19}) which is
interpreted as fluctuations by Padmanabhan and Myung. Let us
consider the solution to this apparent discrepancy. Relation
(\ref{eq17}) is derived from considerations concerning fluctuations
by use of a Taylor expansion. It can be considered in the framework
of renormalization as the truncature of a convergent series up to
its first terms (see \cite{article30} after their Eq. (\ref{eq34})).
In particular, Eq. (\ref{eq17}) is valid for the two following
limits:

For $r\gg r_{1}$, Eq (\ref{eq17}) represents the first terms of a Taylor expansion
and $\Lambda $ tends to (\ref{eq7}) which itself tends to its minimal constant limit $%
\Lambda _{m}=\frac{3}{r_{M}^{2}}$ when $r$ tends to its maximum limit $r_{M}$
(IR cutoff). For these large values of $r$, we are concerned with the inflation undergone today by the universe.
 Therefore the
fluctuations we considered in (\ref{eq13}) and (\ref{eq17}) are these recent fluctuations. In this limit, the model (\ref{eq7})
was studied in \cite{article27}.

For $r\ll r_{1}$, Eq (\ref{eq17}) can be rewritten as a Taylor
expansion for little values of $r$
\begin{equation}
\Lambda =\frac{3r_{1}^{2}}{r^{4}}(1+(\frac{r}{r_{1}})^{2})\text{,}
\label{eq20}
\end{equation}
and $\Lambda$ tends towards another de Sitter state when $r$ reaches
its minimum limit $r_{m}$ (UV cutoff). For these values of $r$, what
is concerned is the  maximum value of $\Lambda $, during the
inflation formerly undergone by the very early universe: $\Lambda _{M}=\frac{%
3r_{1}^{2}}{r_{m}^{4}}$. Padmanabhan and Myung consider these early fluctuations in (\ref{eq18}).

Considered as a renormalisation relation, Eq. (\ref{eq17}) is also valid for any
intermediate value of $r$ and especially when $r\equiv r_{1}$. Assuming the
given expression (\ref{eq17}) for $\Lambda$, we derive now the evolution of the universe.

\section{Evolution equations for H and r}

In a spatially flat FLRW cosmology ($k=0$, metrics signature + - -
-), the field equations (\ref{eq2}) for a two-component perfect
fluid with the variable CC as first component are given
by%
\begin{equation}
(\Lambda +\chi \rho )c^{2}=3H^{2}
\label{eq21}
\end{equation}%
\begin{equation}
\overset{\cdot }{H}\text{ }=-\frac{\gamma }{2}\chi c^{2}\rho \text{
.} \label{eq22}
\end{equation}
From now on, the index 2 for the second component of the fluid is
understood. Equations (\ref{eq21}) and (\ref{eq22}) lead to%
\begin{equation}
\overset{\cdot }{H}\text{ }=-\frac{\gamma }{2}(3H^{2}-\Lambda c^{2})\text{.}
\label{eq23}
\end{equation}
By defining%
\begin{equation}
Y:=\frac{1}{r^{2}}\text{ ,}
\label{eq24}
\end{equation}
(\ref{eq23}) and (\ref{eq10}) become
\begin{equation}
2\overset{\cdot }{H}+3\gamma H^{2}-3\gamma c^{2}Y(1+\alpha Y)=0
\label{eq25}
\end{equation}%
\begin{equation}
\overset{\cdot }{Y}+2HY-2cY^{\frac{3}{2}}=0\text{ ,}
\label{eq26}
\end{equation}
where the expression (\ref{eq17}) of $\Lambda$ has been introduced. Eqs (\ref{eq25}) and
(\ref{eq26}) constitute a system of two coupled first order differential equations
satisfied by \{$Y(t),H(t)$\}. We introduce the dimensionless variables%
\begin{equation}
y:=\frac{Y}{Y_{1}}\text{ , }h:=\frac{H}{H_{1}}\text{ ,}
\label{eq27}
\end{equation}
where $Y_{1}$ and $H_{1}$ are two, \textit{a priori} independent, arbitrary
constant scales for $Y$ and $H$ respectively. We also define the
dimensionless time $\tau $%
\begin{equation}
\tau :=(H_{1}t)^{n}\text{ , (}n> 0\text{).}
\label{eq28}
\end{equation}
Hence, (\ref{eq25}) and (\ref{eq26}) become%
\begin{equation}
2n\tau ^{\frac{n-1}{n}}h^{\prime }+3\gamma h^{2}-3\gamma X_{1}y(1+\alpha
_{1}y)=0
\label{eq29}
\end{equation}%
\begin{equation}
n\tau ^{\frac{n-1}{n}}y^{\prime }+2hy-2\sqrt{X_{1}}y^{\frac{3}{2}}=0\text{ ,}
\label{eq30}
\end{equation}
with
\begin{equation}
X_{1}:=\frac{Y_{1}c^{2}}{H_{1}^{2}}
\label{eq31}
\end{equation}
the dimensionless ratio of the scales $Y_{1}$ and $H_{1}$ and
\begin{equation}
\alpha _{1}:=\alpha Y_{1}\text{.}
\label{eq32}
\end{equation}
The apostrophe denotes the derivative with respect to $\tau $.
Finally, we set
\begin{equation}
\lambda :=\frac{\Lambda }{\Lambda _{1}}\text{ ,}
\label{eq33}
\end{equation}
where $\Lambda _{1}$ is an arbitrary constant scale for $\Lambda .$ With
(\ref{eq17}), (\ref{eq33}) yields%
\begin{equation}
\lambda =\frac{y}{l_{1}}(1+\alpha _{1}y)
\label{eq34}
\end{equation}
where%
\begin{equation}
l_{1}:=\frac{\Lambda _{1}}{3Y_{1}}\text{ ,}
\label{eq35}
\end{equation}
is the dimensionless ratio of the scales $\Lambda _{1}$ and $Y_{1}$.We deduce%
\begin{equation}
\frac{\Lambda _{1}c^{2}}{3H_{1}^{2}}=X_{1}l_{1}
\label{eq36}
\end{equation}
which furnishes a scale for the dimensionless CC density parameter $\Omega
_{\Lambda }$ defined as%
\begin{equation}
\Omega _{\Lambda }:=\frac{\Lambda c^{2}}{3H^{2}}\leq 1\text{ ,}
\label{eq37}
\end{equation}
or, with (\ref{eq33})-(\ref{eq36}) and (\ref{eq27}),%
\begin{equation}
\Omega _{\Lambda }:=X_{1}\frac{y(1+\alpha y)}{h^{2}}\text{ .}
\label{eq38}
\end{equation}
Finally, setting $\tau =0$  $\Omega _{\Lambda }(0)=1$, with (\ref{eq38}) gives
\begin{equation}
h(0)=[X_{1}y(0)(1+\alpha _{1}y(0))]^{\frac{1}{2}}\text{.}
\label{eq39}
\end{equation}
Now it only remains to fix $y(0)$, besides the parameters $X_{1}$, $\alpha
_{1}$, $n$ , $l_{1}$ and $\gamma $.

\section{Results}

We numerically solve the equations (\ref{eq29}) and (\ref{eq30}) and obtain the solutions $%
y(\tau )$ and $h(\tau )$, for the initial conditions (\ref{eq39}) with%
\begin{equation}
y(0)=10^{93}\text{ , }\gamma =1\text{, }l_{1}=1\text{, }X_{1}=10^{-3}\text{,
}\alpha _{1}=10^{-55}\text{, and }n=\frac{1}{220}\text{ .}
\label{eq40}
\end{equation}

By choosing $r_{2}=$ $r_{3}=r_{M}\simeq 10^{26}m$ (see Section 4, Eq. (\ref{eq16})),
the condition (\ref{eq16}) is satisfied for any $r\geq r(0)\simeq 10^{-11}m$ (this
last value will be derived in the following).

The functions $\Omega _{\Lambda }(\tau )$ and $\lambda (\tau )$ are deduced
from (\ref{eq38}) and (\ref{eq34}) respectively and the function $\Omega _{\Lambda }(\tau )$
is plotted in figures 1, 2 and 3 for different ranges of time.

\begin{figure}[ht]
\begin{center}
\includegraphics[scale=0.7]{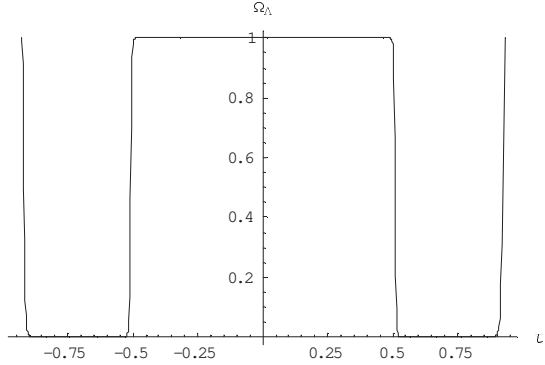}
\end{center}
\caption{Plot of the vacuum density parameter $\Omega_{\Lambda }$ as
a continuous function of dimensionless parameter time $\tau $ for
$\tau \in \left[ -0.930352,0.930352\right] $. The initial inflation
($\tau =0$) and the final de Sitter stage appear. The function
$\Omega_\Lambda(\tau)$ was parametrized by the interpolating
functions $y(\tau)$ and $h(\tau)$, and the numerical solutions of
Eq. (\ref{eq29}) and (\ref{eq30}) with the initial conditions
(\ref{eq39}) and (\ref{eq40}).} \label{figure1}
\end{figure}

\begin{figure}[ht]
\begin{center}
\includegraphics[scale=0.7]{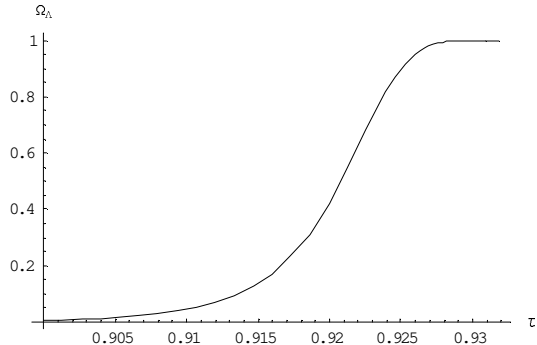}
\end{center}
\caption{Plot of $\Omega_\Lambda$ versus $\tau$ at the end of the
evolution. The universe enters again on a phase of acceleration for
$\Omega_\Lambda >1/3$ and tends to de Sitter (plateau
$\Omega_\Lambda=1$). The present time $\tau_0=0,9220725$ i.e.
$t_0=4,88.10^{17}\,s$ corresponds to $\Omega_{\Lambda 0}=0.63$ (see
table 1).} \label{figure2}
\end{figure}
\begin{figure}[ht]
\begin{center}
\includegraphics[scale=0.7]{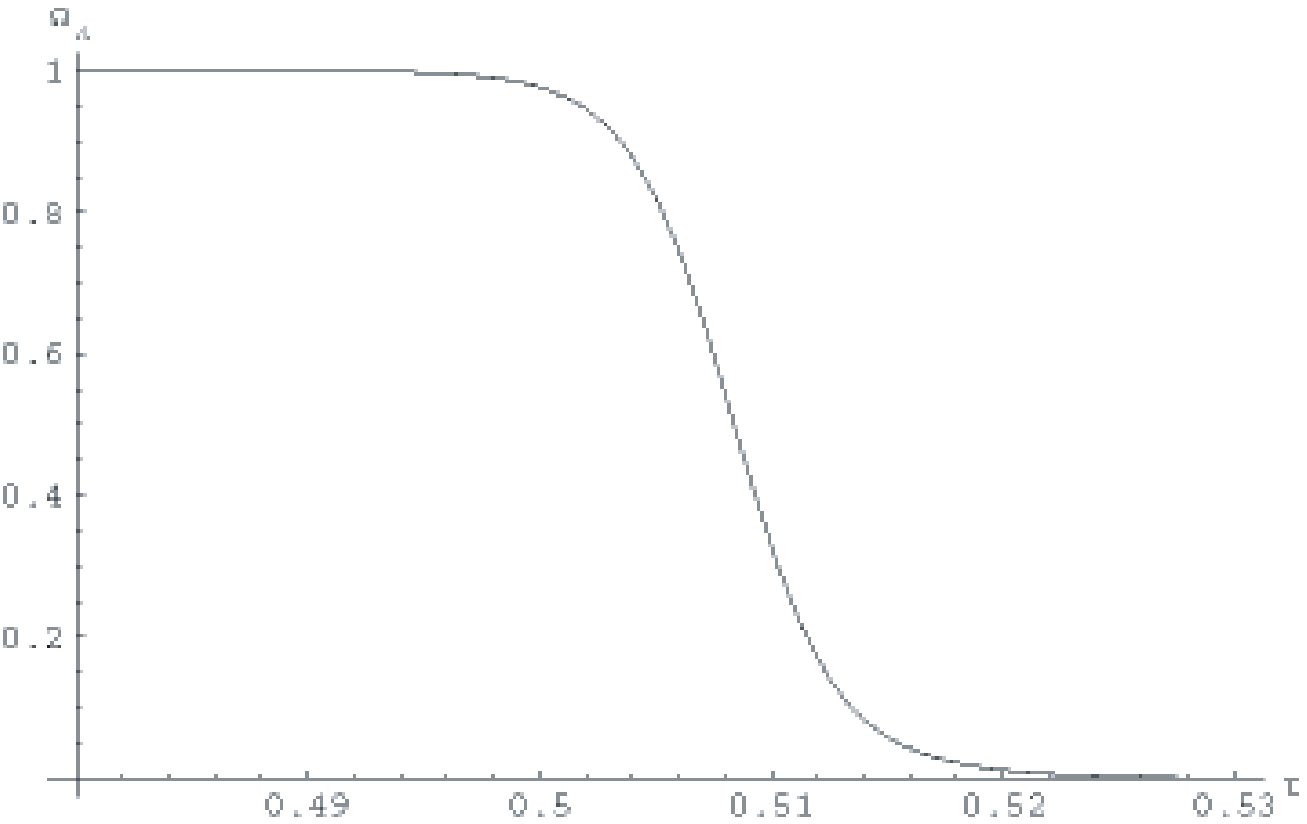}
\end{center}
\caption{Plot of $\Omega_\Lambda$ versus $\tau$ at the beginning of
the evolution. There is an initial inflation (de Sitter
$\Omega_\Lambda=1$), then a graceful exit of inflation for
$\Omega_\Lambda<1/3$ (see table 1).} \label{figure3}
\end{figure}

As the current values of the observational data ($\Omega _{\Lambda }(\tau
_{0})=0.7$ , $H(\tau _{0})=72$ $kms^{-1}Mpc^{-1}$) of the $\Lambda CDM$
model \cite{article33}\cite{article34} are model-dependent, we use the
values derived in the framework of our $\Lambda(t)CDM$ model from the
studies of SnIa observational constraints \cite{article27}%
\begin{equation}
\Omega _{\Lambda }(\tau _{0})=0.63\text{ , }H(\tau _{0})=64\text{ }%
kms^{-1}Mpc^{-1}\text{ ,}
\label{eq41}
\end{equation}
where $\tau _{0}$ represents the present instant. Let us emphasize that the
use of the values derived from the $\Lambda CDM$ concordance model would
only marginally modify the following results.

The different scales $H_{1}$,$Y_{1}$, and $\Lambda _{1}$ can be deduced from these two values (\ref{eq41}). We deduce the instant $\tau _{0}$ from the plot of the
function $\Omega (\tau )$
\begin{equation}
\tau _{0}=0.9220725\text{ .}
\label{eq42}
\end{equation}
Setting this value in $y(\tau )$ and $h(\tau )$ and using (\ref{eq34}), we deduce
\begin{equation}
y(\tau _{0})=2.0576\times 10^{18}\text{ , }h(\tau _{0})=5.7149\times 10^{7}%
\text{ , }\lambda (\tau _{0})=2.0576\times 10^{18}\text{ .}
\label{eq43}
\end{equation}
Then, Eq. (\ref{eq27}) with the numerical values (\ref{eq41}) and (\ref{eq43}) yields%
\begin{equation}
H_{1}=\frac{H(\tau _{0})}{h(\tau _{0})}=3.62928\times 10^{-26}s^{-1}\text{ .}
\label{eq44}
\end{equation}
From (\ref{eq31}), (\ref{eq40}) and (\ref{eq44}), we have%
\begin{equation}
Y_{1}=\frac{X_{1}H_{1}^{2}}{c^{2}}=1.46555\times 10^{-71}m^{-2}\text{.}
\label{eq45}
\end{equation}
Besides, $\Lambda (\tau _{0})$ is directly fixed by the observational data
(\ref{eq41}) from (\ref{eq37})
\begin{equation}
\Lambda (\tau _{0})=\frac{3\Omega _{\Lambda }(\tau _{0})H^{2}(\tau _{0})}{%
c^{2}}=9.04649\times 10^{-53}m^{-2}\text{ .}
\label{eq46}
\end{equation}
Thus with (\ref{eq33}) and (\ref{eq43})
\begin{equation}
\Lambda _{1}=\frac{\Lambda (\tau _{0})}{\lambda (\tau _{0})}=4.39662\times
10^{-71}m^{-2}\text{.}
\label{eq47}
\end{equation}
The numerical value of $r_{1}$ linked to the scale $Y_{1}$ and identified to
the value $L_{1}$ of Eq. (\ref{eq19}) can now be estimated from (\ref{eq32}), (\ref{eq40}) and (\ref{eq45}):
$r_{1}=8.26037\times 10^{7}m$. It is an intermediary scale between $%
L_{P}\simeq 8.02121\times 10^{-35}m$ (for $l_{P}^{\prime }:=\sqrt{\frac{%
\text{%
h{\hskip-.2em}\llap{\protect\rule[1.1ex]{.325em}{.1ex}}{\hskip.2em}%
}G}{c^{3}}}\simeq 1.6\times 10^{-35}m$, with $L_{P}=2\sqrt{2\pi }%
l_{P}^{\prime }$ ) and $L_{\Lambda today}=r(\tau _{0})\simeq 1.82104\times
10^{26}m$.
We then obtain
\begin{equation}
\frac{\Lambda (0)}{\Lambda (\tau _{0})}=\frac{y(0)(1+\alpha _{1}y(0))}{%
y(\tau _{0})(1+\alpha _{1}y(\tau _{0}))}\simeq 4.86003\times 10^{112}\text{ .%
}
\label{eq48}
\end{equation}
The dark energy density parameter $\Omega _{\Lambda }(\tau )$ decreases from
the initial value $\Omega _{\Lambda }(0)=1$ and then reaches the value $1/3$%
, i.e. the end of the acceleration (or equivalently the exit of inflation),
which occurs at the instant $t_{2}=1.23824\times 10^{-39}s$ (evaluated from
Eq. (\ref{eq28})).

Then, it passes through a minimum $\Omega _{\Lambda m}(\tau
_{m})=4.99970\times 10^{-20}$ at the instant $t_{m}=2.9042\times 10^{-11}s$
where $\Lambda (\tau _{m})=8.795\times 10^{-16}m^{-2}$ and $H(\tau
_{m})=2.2955\times 10^{10}s^{-1}$.
After, $\Omega _{\Lambda }(\tau )$ increases again ( $\Lambda $ comes back
[Krauss and Turner 1995 34b]) until a return of the inflation which occurs
at the instant $t_{3}=1.81\times 10^{17}s=5.73\, Gyr$ .

It reaches its present day value $0.63$ at $\tau _{0}$ (\ref{eq42}) or $%
t_{0}=4.88084\times 10^{17}s=15.4664Gyr$. This value is in good agreement
with the observational results derived from the ``concordance model" (\cite%
{article34} Section 2.5. p.3) and allows to avoid the age crisis (see e.g.
\cite{article34c}). With a linear extrapolation of $H(\tau )$ during $\tau
_{3}-\tau _{0}$, we roughly evaluate the redshift, namely $z(\tau
_{3})\simeq 1.01826$, where the acceleration begins again. This value is in
good agreement up to the previous approximation with the exact value of the
transition redshift $z_{T}=0.965$ found in \cite{article27}.

In the same time, the ``coincidence problem'' is resolved. The date
of the coincidence can be defined by different ways : for instance
by the instant $t_{m}$ (see above), i.e. the instant from which
$\Omega _{\Lambda }$ grows again, or by the instant $t_{3}$ (see
above) when the universe inflates again, or even, more recently, by
the instant $t_{4}=3.61505\times 10^{17}s=11.4554Gyr$ when $\Omega
_{\Lambda }=\frac{1}{2}=\Omega _{m}$. In the following, we use this
last definition.

If we admit that the recombination starts at the instant $t_{\ast
}=1.17\times 10^{13}s=372kyr$, and ends at $t_{\ast }^{\prime
}=1.54\times 10^{13}s=487kyr$ , the values of the main cosmological
parameters are, for these instants : $\Omega _{\Lambda }(\tau _{\ast
})=1.43\times 10^{-4}$, $\Omega _{\Lambda }(\tau _{\ast }^{^{\prime
}})=1.722\times 10^{-4}$, $H(\tau _{\ast })=5.68\times 10^{-14}s^{-1}$, $%
H(\tau _{\ast }^{\prime })=4.34\times 10^{-14}s^{-1}$, $\Lambda (\tau _{\ast
})=1.54\times 10^{-47}m^{-2}$, $\Lambda (\tau _{\ast }^{\prime })=1.08\times
10^{-47}m^{-2}$. A determination of the value of $\Omega _{\Lambda }(\tau
_{\ast }^{^{\prime }})$ ( or of $\Omega _{m}(\tau _{\ast }^{^{\prime }})$ )
from observational data, if feasible, would be an interesting test for the
model. The following table presents the main results of the FLRW e.h. model.

\begin{tabular}[t]{|c|c|c|c|c|}
\hline  $\tau $ & $t(s)$ & $H(s^{-1})$ & $\Omega _{\Lambda }$ &
$\Lambda (m^{-2})$\vspace*{0.1cm}\\
 \hline \hline
 $0$ & $0$ & $4.574$ $10^{38}$ & $1$ &
$6.98$ $10^{60}$
 \vspace*{0.1cm}\\
  \hline  $\tau _{2}=0.50993$ & $t_{2}=1.24$ $10^{-39}$
& $2.79$ $10^{38}$ & $0.333$ & $8.685$ $10^{59}$
\vspace{0.1cm}\\
 \hline
$\tau _{m}=0.6862$ & $t_{m}=2.90$ $10^{-11}$ & $2.29$ $10^{10}$ & $5$ $%
10^{-20}$ & $8.795$ $10^{-16}$  \vspace*{0.1cm} \\
\hline
$\tau _{\ast }=0.8786$ & $t_{\ast }=1.17$ $10^{13}$ & $5.68$ $10^{-14}$ & $%
1.43$ $10^{-4}$ & $1.542$ $10^{-47}$  \vspace*{0.1cm}\\
\hline
$\tau _{\ast }^{\prime }=0.8796$ & $t_{\ast }^{\prime }=1.537$ $10^{13}$ & $%
4.34$ $10^{-14}$ & $1.72$ $10^{-4}$ & $1.082$ $10^{-47}$
\vspace*{0.1cm} \\
\hline
$\tau _{3}=0.9189$ & $t_{3}=2.29$ $10^{17}$ & $3.52$ $10^{-18}$ & $0.333$ & $%
1.378$ $10^{-52}$  \vspace*{0.1cm} \\
\hline
$\tau _{4}=0.92081$ & $t_{4}=2.87$ $10^{17}$ & $3.15$ $10^{-18}$ & $0.5$ & $%
1.655$ $10^{-52}$  \vspace*{0.1cm} \\
\hline
$\tau _{0}=0.92207$ & $t_{0}=4.88$ $10^{17}$ & $2.07$ $10^{-18}$ & $0.63$ & $%
9.046$ $10^{-53}$ \vspace*{0.1cm}  \\
\hline
\end{tabular}
\vspace*{0.2cm}\\
Table1. Values of $H$ , $\Omega _{\Lambda }$ and $\Lambda $ at some
key
times of the evolution.\\

The long period about from $6.39\times 10^{-40}s$ to $t_{4}\simeq 2.8\times
10^{17}s$ when $\Lambda $ does not dominate, i.e. when $\Omega _{\Lambda }< 0.5
$, allows to maintain the standard model and its afferent results all along
this period.

Another interesting result to be noted is the shape of the curve $\Omega
_{\Lambda }(\tau )$ ( fig.1) which presents a symmetry with respect to $\tau
=0$ , suggesting a period for $\Omega _{\Lambda }(\tau )$. Each curve ( $%
\Omega _{\Lambda },h,y,\lambda ,H,\Lambda $) is (quasi-) symmetrical with
respect to $t=0$. The functions $\Lambda (\tau )$ and $H(t)$ decrease
(increase resp.) when $\tau \geq 0$ ($\tau \leq 0$ resp.). There is a ``big
bounce" on an inflationary space at $t=0$, with a period that is stretched
over more than twice $t_{0}$ and another one at a very large time on a
quasi- de Sitter space. We studied above the results given for $\tau \geq 0$
only. The first inflation is ``strong" (at $t=0$) when $\Lambda $ is
maximum, the second one is ``weak"(at $t\rightarrow \infty $) when $\Lambda $
is minimum (possibly$\rightarrow 0$). Today, $\Lambda $ is on the way to its
minimum (maybe zero), and the universe is dominated by the small
cosmological constant (\ref{eq36}).

We can also obtain the $R(t)$ ( or ``jerk" $j(t)$ ) and $S(t)$
``statefinder" parameters \cite{article35} and $q(t)$ deceleration
parameter for the model
\begin{equation}
q\equiv \frac{-\overset{\cdot \cdot }{a}}{aH^{2}}=-1+\frac{3\gamma }{2}%
(1-\Omega _{\Lambda })
\label{eq49}
\end{equation}%
\begin{equation}
R\equiv \frac{\overset{\cdot \cdot \cdot }{a}}{aH^{3}}=3X_{1}\frac{%
y(1+2\alpha _{1}y)}{h^{3}}(\sqrt{X_{1}y}-h)+1
\label{eq50}
\end{equation}%
\begin{equation}
S\equiv \frac{R-1}{3(q-\frac{1}{2})}\text{ ,}
\label{eq51}
\end{equation}%
with for their present day values
\begin{equation}
q_{0}=-0.445\text{ , }R_{0}=0.610125\text{, }S_{0}=0.137522\text{ .}
\label{eq52}
\end{equation}

The positive sign of $R_{0}$ indicates the present overinflation, (either
increasing of inflation, or trend towards inflation). One can also evaluate $%
R$ in the early period of deflation (i.e. desinflation). For instance at $%
\tau =0.509275$ when inflation was present, though decreasing, because $%
\Omega _{\Lambda }(\tau =0.509275)=0.4> \frac{1}{3}$, we obtain $R(\tau
=0.509275)=-1.414< 0$.We can yet obtain $R< 0$ for $\Omega _{\Lambda }< \frac{1}{3%
}$ during some time after the end of the initial inflation. For example at $%
\tau =0.51151$ when $\Omega _{\Lambda }=0.2<\frac{1}{3}$, we have yet $%
R(\tau =0.51151)=-0.20005$. The instant when $R$ changes of sign is given by
$\tau _{9}=0.51204075$ or $t_{9}=3.07026\times 10^{-39}s$. Since then, $R$
is positive.

The comparison of the ``kinematical" values (\ref{eq52}) to observations (SNAP
supernova data ) could also be a test for our model (see the remark of \cite%
{article35}, Section 4) and would allow the measurement of the production
source of CDM \cite{article36} which is constrained by the statefinder
parameters and the ``snap" parameter $s:=\frac{\overset{(4)}{a}}{aH^{4}}$.

Finally, this model also supports the low-$l$ CMBR power spectrum
suppression suspected in the WMAP data. Indeed, the event horizon plays
naturally the role of a dynamical IR cut-off translatable to a wave
number cutoff \cite{article37} \cite{article38} as for the holographic model
\cite{article21} \cite{article26}. Applying the same procedure as \cite%
{article37} \cite{article38} for the Dirichlet boundary condition
$\lambda _{c}=2r_{0}$, we roughly evaluate the influence of this
cutoff. In the integrated Sachs-Wolfe (ISW) effect, the IR cutoff
appears as the lower limit of the comoving momentum which truncates
the summation giving the coefficient $C_{l}$ of the power spectrum.
For a power law spectrum with a
spectral index $n=1$ (Harrisson-Zeldovich), we have for the low $l$ \cite%
{article37}:%
\begin{equation}
C_{l}\simeq N\sum_{k> k_{c}}\frac{1}{k}j_{l}^{2}(k\eta _{0})
\label{eq53}
\end{equation}
where $j_{l}$ are the spherical Bessel functions. In flat spacetime, the
multipole number $l$ is given by
\begin{equation}
l=k_{l}(\eta _{0}-\eta ^{\ast })
\label{eq54}
\end{equation}
where $(\eta _{0}-\eta ^{\ast })$ is the comoving distance to the last
scattering surface.The definition of the conformal time in flat space $\eta
=-\frac{r}{a}$ with the Dirichlet limits condition leads to%
\begin{equation}
l_{c}=\pi ((1+z^{\ast })\frac{r^{\ast }}{r_{0}}-1)\text{.}
\label{eq55}
\end{equation}

Taking the current value $z^{\ast }=1089$ \cite{article5,article5b} and
using (\ref{eq17}), we obtain
\begin{equation}
r=\sqrt{\frac{3}{2\Lambda }(1+\sqrt{1+\frac{4}{3}\alpha \Lambda })}\text{ .}
\label{eq56}
\end{equation}
$\frac{r^{\ast }}{r_{0}}$ can be evaluated with $\Lambda ^{\ast }$and $%
\Lambda _{0}$ from Table 1 and we find : $l_{c}=5.15$. For the value $%
\Lambda ^{\prime \ast }$ of the end of the recombinaison, we get : $%
l_{c}^{\prime }=6.76$. These values are in the expected range.

\section{Conclusion}

From the FLRW e.h. entropy with thermal fluctuations (Eq.
(\ref{eq13})), we obtained the corresponding law for the
cosmological constant $\Lambda (r)$ as a function of the event
horizon radius $r$ (Eq. (\ref{eq17})). We used the striking
similarity of this last expression (\ref{eq17}) with a
renormalization relation (Eq. (\ref{eq19})) to reinterpret
(\ref{eq17}) in terms of such a (truncated) renormalization relation
(\ref{eq19}).

Using this $\Lambda (r)$ law in an interacting two-component
$\Lambda (t)CDM$ cosmological model in a FLRW spacetime, we derived
a system of coupled equations for $H(t)$ and $r(t)$ (Eqs.
(\ref{eq29}) and (\ref{eq30})). Through a numerical resolution
(Section 6), we obtained the curves of evolution of $\Omega
_{\Lambda }$ (figs. 1 to 3), the table 1 of the values of the main
cosmological parameters ($H,\Omega _{\Lambda }$) at different times
in the evolution of the universe, and the evolution of the
deceleration and statefinder parameters.

The event horizon thermodynamical model presented here allows us to account
for most of essential features of the cosmology such as the value of the age
of the universe,$\sim 15.5Gyr$ (safe from the age problem \cite{article34b}%
), the ratio $\frac{\Lambda \lbrack 0]}{\Lambda \lbrack \tau _{0}]}\sim
5\times 10^{112}$, as well as the exit of inflation after about $t_{2}\simeq
23000$ Planck times, the existence of a minimum $\sim 5$ $\times 10^{-20}$
for $\Omega _{\Lambda }$ attained after about $3$ $\times 10^{-11}s$, the
coming back of the inflation (after about $7.24Gyr$ from the beginning),
followed by the recent coincidence, at $11.4Gyr$, with the matter density,
but also allows us to address the entropy problem, and the creation of
(dark) matter from the vacuum energy. Besides, the model does not perturb
the results of the old standard model for the period going from $6.39\times
10^{-40}s$$\simeq 10000$ Planck times to $2.87\times 10^{17}s=11.4Gyr$ all
along which the DE is subdominant ( $\Omega _{\Lambda }<0.5$ ). Finally, it
can render an account of the low-$l$ CMBR power spectrum suppression, for an
angular scale corresponding to multipole numbers lower than $l_{c}\sim 7$.

\end{document}